 \documentstyle[prl,aps,multicol,epsfig]{revtex}
\begin{document}
\draft \date{\today} \title
  {Superconductors in realistic geometries: Geometric edge barrier
   versus pinning}
\author{Ernst Helmut Brandt}
\address{Max-Planck-Institut f\"ur Metallforschung,
   D-70506 Stuttgart, Germany}
\maketitle
  \begin{center}  \vspace{-3.1 cm} {\small For Physica C:~
  Euro-Conference on Vortex Matter in Superconductors, Crete,
  18-24 September 1999} \vspace{2.1 cm} \end{center}

\begin{abstract}
   The magnetic response of type-II superconductors can be
irreversible due to two different reasons: vortex pinning and
barriers for flux penetration. Even without bulk pinning and
in absence of a microscopic Bean-Lingston surface barrier for
vortex penetration, superconductors of nonellipsoidal shape
can exhibit a large geometric barrier for flux penetration.
This edge barrier and the resulting irreversible magnetization
loops and flux-density profiles
are computed from continuum electrodynamics for superconductor
strips and disks with constant thickness, both without and with
bulk pinning. Expressions for the field of first flux entry
$H_{\rm en}$ and for the reversibility field $H_{\rm rev}$
above which the pin-free magnetization becomes
reversible are given. Both fields are proportional to the lower
critical field $H_{c1}$ and else depend only on the specimen shape.
These realistic results are compared with the reversible magnetic
behavior of ideal superconductor ellipsoids.
\end{abstract}
    \begin{multicols}{2}
    \narrowtext

\section{Introduction}  

  The irreversible magnetic behavior of type-II superconductors
usually is caused by pinning of the vortices at inhomogeneities
in the material \cite{1}. However, similar hysteresis
effects were also observed \cite{2} in type-I superconductors,
which do not contain flux lines, and in type-II superconductors
with negligible pinning. In these two cases the
magnetic irreversibility is caused by a geometric (specimen-shape
dependent) barrier which delays the penetration of magnetic flux
but not its exit. In this respect the {\it macroscopic} geometric
barrier behaves similar as the {\it microscopic} Bean-Livingston
barrier \cite{3} for straight vortices penetrating at a parallel
surface. In both cases the magnetic irreversibility is caused by
the asymmetry between flux penetration and exit.
The geometric irreversibility is most pronounced for thin
films of constant thickness in a perpendicular field. It is absent
only when the superconductor is of exactly ellipsoidal shape or
is tapered like a wedge with a sharp edge where flux can penetrate
easily due to the large local enhancement of the external magnetic
field at this edge in a diamagnetic material.

   Ellipsoids are a particular case. In superconducting ellipsoids
the inward directed driving force exerted on the vortex ends by the
surface screening currents is exactly compensated by the vortex line
tension \cite{4}. An isolated vortex line is thus in an indifferent
equilibrium at any distance from the specimen center. The repulsive
vortex interaction therefore yields a uniform flux density and the
magnetization is reversible. However, in specimens with
constant thickness (i.e.\ with rectangular cross-section) this line
tension opposes the penetration of flux lines at the four corner
lines, thus causing an edge barrier; but as soon as two penetrating
vortex segments join at the equator they contract and are driven to
the specimen center by the surface currents, see Figs.\ 1 and 2. As
opposed to this, when the specimen profile is tapered and has a
sharp edge, the driving force of the screening currents even in
very weak applied field exceeds the restoring force of the line
tension such that there is no edge barrier. The resulting absence
of hysteresis in wedge-shaped samples was clearly shown by
Morozov et al.\ \cite{5}.

   For thin superconductor strips with an edge barrier an elegant
analytical theory of the field and current profiles has been
presented by Zeldov et al.\ \cite{6}, using the theory of complex
functions, see also the calculations \cite{7,8}.
With increasing applied field $H_a$, the magnetic flux does not
penetrate until an entry field $H_{\rm en}$ is reached;
at $H_a = H_{\rm en}$ the flux immediately jumps to
the center, from where it gradually fills the entire strip or
disk. This behavior in increasing $H_a$ is similar to that of
thin films with artificially enhanced pinning near the
edges \cite{7,9}, but in decreasing $H_a$ the behavior is
different: In films with enhanced edge pinning (critical current
density $J_c^{\rm edge}$) the current density $J$ at the edge
immediately jumps from $+J_c^{\rm edge}$ to  $-J_c^{\rm edge}$
when the ramp rate reverses its sign, while in pin-free
films with geometric barrier the current density at the edge
first stays constant or even increases and then
gradually decreases and reaches zero at $H_a=0$. For pin-free
thin strips the entry field $H_{\rm en}$ was estimated
in Refs.\ \cite{6,10,11}.

  The outline of the present work is as follows. Section 2
discusses the reversible magnetic behavior of pin-free superconductor
ellipsoids. The effective demagnetization factor of long strips
(or slabs) and circular disks (or cylinders)  with rectangular
cross section $2a \times 2b$ is given in Sec.\ 3. In Sec.\ 4
appropriate continuum equations and algorithms are presented that
allow to compute the magnetic irreversibility caused by pinning and/or
by the geometric barrier in type-II superconductors of arbitrary
shape, in particular of strips and disks with finite thickness.
Results for thick
long strips and disks or cylinders with arbitrary aspect ratio
$b/a$ are given in Sec.\ 5 for
  \linebreak  \vspace{-.65cm} 
  \begin{figure}[F1]
\epsfxsize= .7\hsize  
\centerline{ \epsffile{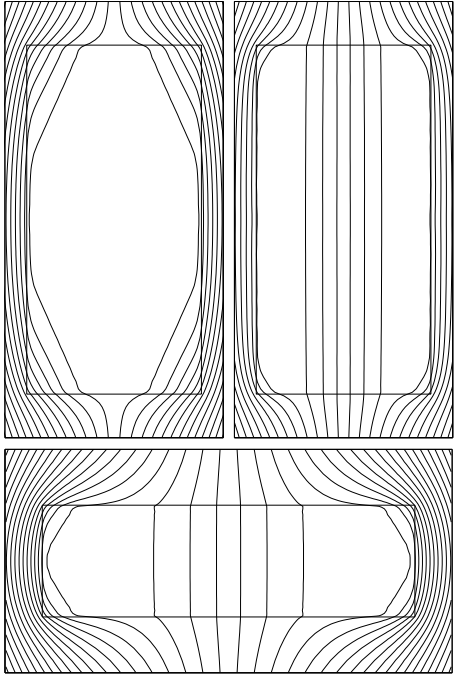}} \vskip .5\baselineskip
\caption{Field lines of the induction ${\bf B}(x,y)$ in strips
 with aspect ratio $b/a=2$ (top) and $b/a =0.3$ (bottom) in
 perpendicular magnetic field $H_a$.  Top left: $H_a/H_{c1} = 0.66$,
 in increasing field shortly before the entry field
 $H_{\rm en}/H_{c1}=0.665$. Top right: $H_a/H_{c1} = 0.5$, decreasing
 field.  Bottom: $H_a/H_{c1} = 0.34$ in increasing field just
 above $H_{\rm en}/H_{c1}= 0.32$. Note the nearly straight field lines
 in the corners indicating the tension of the flux lines.
 The field lines of cylinders look very similar.
 } \end{figure}

\noindent
pin-free superconductors and in
Sec.\ 6 for superconductors with arbitrary bulk pinning.
In particular, explicit expressions are given for the field
of first flux entry $H_{\rm en}$ and for the reversibility field
$H_{\rm rev}$ above which the magnetization curve is reversible
and coincides with that of an ellipsoid.
\section{Ellipsoids}  

   First consider the known magnetization of ideal ellipsoids. If
the superconductor is homogeneous and isotropic, the magnetization
curves of ellipsoids $M(H_a; N)$ are {\it reversible} and may be
characterized by a demagnetizing factor $N$. If $H_a$ is along one
of the three principal axes of the ellipsoid then $N$ is a scalar
with $0 \le N \le 1$.  One has $N=0$ for long specimens
in parallel field, $N=1$ for thin films in perpendicular field,
$ N=1/2$ for transverse circular cylinders, and
$N=1/3$ for spheres. For general ellipsoids with semi-axes
$a$, $b$, $c$ along the cartesian axes $x$, $y$, $z$, the three
demagnetizing factors along the principal axes satisfy
$N_x +N_y +N_z =1$. For rotational ellipsoids with $a=b$ one
has  $N_x=N_y= (1-N_z)/2$ where for cigars with
$a=b < c$ and for disks with $a=b >c$ with eccentricity
$e = |1-c^2/a^2|^{1/2}$ one obtains  \cite{12}
  \begin{eqnarray}  
  N_z &=& {1-e^2 \over e^3} ({\rm atanh}\,e -e),~~\mbox{(cigar),}
                            \nonumber \\
  N_z &=& {1-e^2 \over e^3} (e -{\rm atan}\, e),~~~~\mbox{(disk).}
  \end{eqnarray}
For thin ellipsoidal disks with $a \ge b\gg c$  one has \cite{13}
  \begin{eqnarray}  
  N_z = 1 - {c \over b} E(k) \,,
  \end{eqnarray}
where E(k) is the complete elliptic integral of the second kind
with $k^2= 1 -b^2 / a^2$.

  When the magnetization curve in parallel field is
known, $M(H_a; 0) = B/\mu_0 -H_a$ where $B$ is the flux density
inside the ellipsoid, then the homogeneous
magnetization of the general ellipsoid, $M(H_a; N)$, follows
from the implicit equation
  \begin{eqnarray}    
    H_i = H_a - N\, M(H_i; 0)\,.
  \end{eqnarray}

  \begin{figure}[F2]
\epsfxsize= .95\hsize  \vskip 0.5\baselineskip
\centerline{ \epsffile{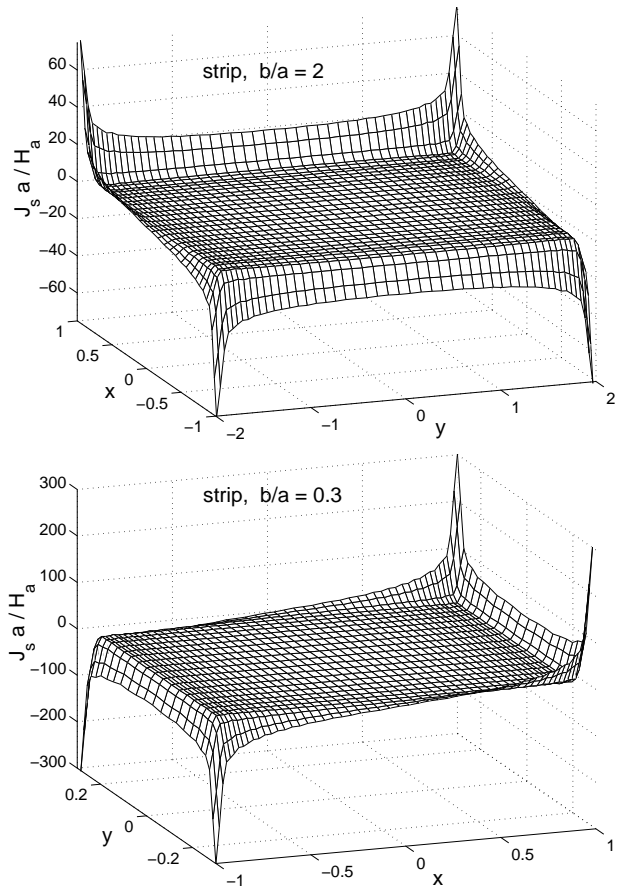}}
\caption{ 3D plots of the screening current density $J_s(x,y)$,
Eq.\ (11), in superconductor strips with $b/a=2$ (top) and $b/a=0.3$
(bottom) as in Fig.\ 1. Shown is the limit of small applied field
$H_a \ll H_{c1}$ before magnetic flux has penetrated.
For better presentation the depicted $J_s(x,y)$ is smeared over a
few grid cells.
  } \end{figure}

\noindent
Solving Eq.\ (3) for the effective internal field $H_i$, one
obtains $M=M(H_a; N) = M(H_i; 0)$. In particular, for
the Meissner state ($B \equiv 0$) one finds $M(H_a;0) = -H_a$ and
  \begin{eqnarray}    
   M(H_a; N) = - {H_a \over 1-N} ~~~{\rm for}~~
               |H_a| \le (1-N)H_{c1} \,.
  \end{eqnarray}
At the lower critical field $H_{c1}$ one has $H_i = H_{c1}$,
$H_a = H_{c1}' = (1-N) H_{c1}$, $B=0$, and $M=-H_{c1}$. Near the
upper critical field $H_{c2}$ one has an approximately linear
$M(H_a; 0) =  \gamma (H_a -H_{c2}) <0$ with $\gamma >0$,
yielding
  \begin{eqnarray}    
    M(H_a; N) = {\gamma \over 1 + \gamma N } (H_a - H_{c2})
   ~~~{\rm for}~~ H_a \approx H_{c2} \,.
  \end{eqnarray}
Thus, if the slope $\gamma \ll 1$ is small (and in general, if
$|M/H_a| \ll 1$ is small), demagnetization effects may be
disregarded and one has $M(H_a; N) \approx M(H_a; 0)$.

   The ideal magnetization curve of type-II superconductors with
$N=0$, $M(H_a; 0)$ or $B(H_a;0)/\mu_0 =H_a +M(H_a;0)$, may be
calculated from Ginzburg-Landau (GL) theory \cite{14}, but to
illustrate the geometric barrier any other model curve may be used
provided $M(H_a; 0) = -M(-H_a; 0)$ has a vertical slope at
$H_a =H_{c1}$ and decreases monotonically in size for $H_a >H_{c1}$.
Below for simplicity I shall assume $H_{c1} \ll H_{c2}$ (i.e.\ large
GL parameter $\kappa \gg 1$) and $H_a \ll H_{c2}$. In this case one
may use the model $M(H_a; 0)= -H_a$  for $|H_a| \le H_{c1}$ and
  \begin{eqnarray}    
    M(H_a; 0) = (H_a/|H_a|) (|H_a|^3 - H_{c1}^3 )^{1/3} - H_a
  \end{eqnarray}
for $|H_a| > H_{c1}$, which well approximates the pin-free
GL magnetization \cite{14}.

\section{Thick strips and disks in the Meissner state}   

In nonellipsoidal superconductors the induction ${\bf B(r})$ in
general is not homogeneous, and so the concept of a demagnetizing
factor does not work. However, when the magnetic moment
${\bf m} = {1\over 2} \int {\bf r \times J(r)} d^3 r$ is directed
along $H_a$, one may define an {\it effective demagnetizing factor}
$N$ which in the Meissner state ($B\equiv 0$) yields the same
slope $M/H_a = -1/(1-N)$, Eq.\ (2), as an ellipsoid with this $N$.
Here the definition $M=m/V$  with $m= {\bf m H}_a/H_a$ and
specimen volume $V$ is used. In particular, for long strips or
slabs and circular disks or cylinders with rectangular
cross-section $2a \times 2b$ in a perpendicular or axial magnetic
field along the thickness $2b$, approximate expressions for the
slopes $M/H_a = m/(VH_a)$ are given in Refs.\ \cite{15,16}.
Using this and defining $q \equiv (|M/H_a| -1)(b/a)$, one obtains
the effective $N$ for any aspect ratio $b/a$ in the form
  \begin{eqnarray}    
  N &=& 1 - 1/(1 + q a/b) \,, \nonumber \\
  q_{\rm strip} &=&  {\pi \over 4 } +0.64
   \tanh\Big[ 0.64{b\over a} \ln \Big( 1.7+1.2{a\over b} \Big) \Big]
    \,, \nonumber \\
  q_{\rm disk}  &=&  {4 \over 3\pi } +{2 \over 3\pi }
   \tanh\Big[ 1.27{b\over a} \ln \Big(1 +{a\over b} \Big) \Big] \,.
       \end{eqnarray}

  \begin{figure}[F3]
\epsfxsize= .95\hsize  \vskip 0.5\baselineskip
\centerline{ \epsffile{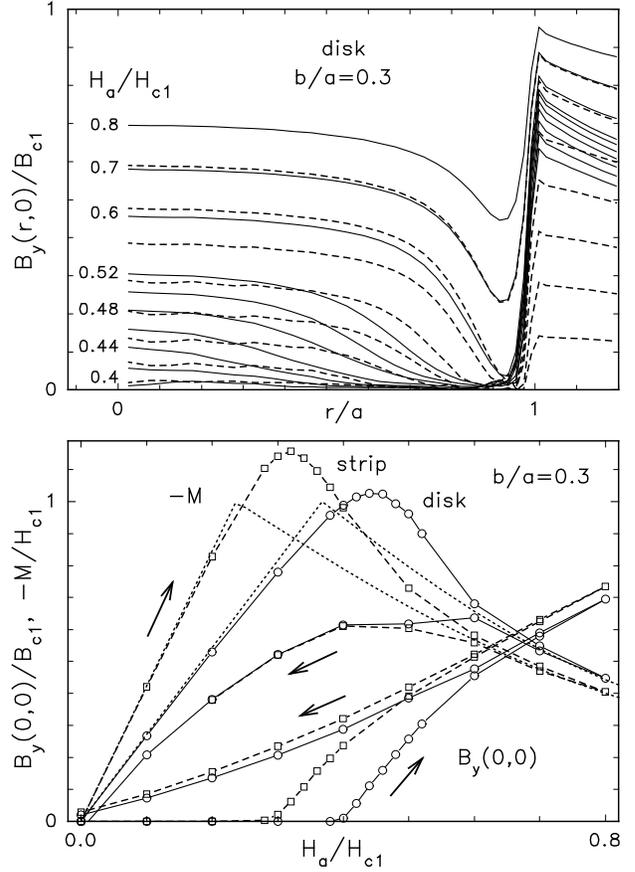}}
\caption{Top: The axial magnetic induction $B_y(r,y)$ in the midplane
 $y=0$ of a pin-free superconductor disk with aspect ratio
 $b/a=0.3$ in increasing field (solid lines) and then decreasing
 field (dashed lines), plotted at
 $H_a/H_{c1}$ = 0.4, 0.42, $\dots$, 0.5, 0.52, 0.6,
 0.7, 0.8, 0.7, 0.6, $\dots$, 0.1, 0.
 Bottom: The induction $B_y(0,0)$ in the center of the same disk
 (solid line) and of a strip (dashed line), both with $b/a=0.3$.
 The symbols mark the field values at which the profiles are
 taken. Also shown is the magnetization loop for the same
 disk and strip and the corresponding reversible magnetization
 (dotted lines).
  } \end{figure}

\noindent
In the limits $b \ll a$ and $b \gg a$, these formulae are exact,
and for general $b/a$ the relative error is $< 1\%$.
For $a=b$ (square cross-section) they yield for the strip $N = 0.538$
(while $N=1/2$ for a circular cylinder in perpendicular field) and for
the short cylinder  $N = 0.365$ (while $N=1/3$ for the sphere).

\section{Computational method}  

    To obtain the full, irreversible magnetization curves $M(H_a)$
of nonellipsoidal superconductors one has to resort to numerics.
Appropriate continuum equations and
algorithms have been proposed recently by Labusch and
Doyle \cite{17} and by the author \cite{18}, based on the Maxwell
\linebreak \pagebreak

  \begin{figure}[F4]
\epsfxsize= .95\hsize  \vskip 0.5\baselineskip
\centerline{ \epsffile{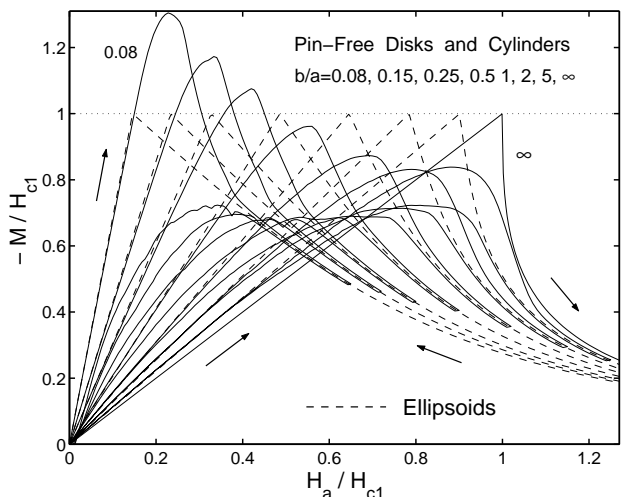}}
\caption{The irreversible magnetization curves $-M(H_a)$ of
 pin-free circular disks and cylinders with aspect ratios
 $b/a=$ 0.08, 0.15, 0.25, 0.5, 1, 2, 5, and $\infty$ in an axial
 field  (solid lines). Here the irreversibility is due only to a
 geometric edge barrier for flux penetration. The reversible
 magnetization curves of the corresponding ellipsoids defined by
 Eqs.\ (3,6,7) are shown as dashed lines.
  } \end{figure}

\noindent
equations and on constitutive laws which describe flux
flow and pinning or thermal
depinning, and the equilibrium magnetization in absence of pinning,
$M(H_a;0)$. For arbitrary specimen shape these two methods proceed
as follows.

   While method \cite{17} considers a magnetic charge density on the
specimen surface which causes an effective field ${\bf H}_i({\bf r})$
inside the superconductor, our method \cite{18} couples the
arbitrarily shaped superconductor to the external field
${\bf B(r},t)$ via surface screening currents: In a first step
the vector potential ${\bf A(r},t)$ is calculated for given current
density ${\bf J}$; then this linear relation (a matrix) is inverted
to obtain ${\bf J}$ for given ${\bf A}$ and given ${\bf H}_a$;
next the induction law is used to obtain the electric field [in our
symmetric geometry one has
${\bf E(J,B)} = -\partial {\bf A}/\partial t$\,], and finally the
constitutive law ${\bf E=E(J,B)}$ is used to eliminate
${\bf A}$ and ${\bf E}$ and
obtain one single integral equation for ${\bf J(r},t)$ as
a function of ${\bf H}_a(t)$, without having to compute
${\bf B(r},t)$ outside the specimen. This method in general
is fast and elegant; but so far the algorithm is restricted
to aspect ratios $0.03 \le b/a \le 30$, and to a number
of grid points not exceeding 1400 (on a Personal Computer).
Improved accuracy is expected by combining the methods \cite{17}
(working best for small $b/a$) and \cite{18}. Here I shall use
the method \cite{18} and simplify it to the two-dimensional
(2D) geometry of thick strips and disks.

 In the 2D geometry of thick strips \cite{15} or short
cylinders \cite{16} in an applied magnetic field
${\bf B_a} =\mu_0{\bf H_a} =\nabla \times {\bf A_a}$ along $y$,
one writes ${\bf r} = (x,y)$ or ${\bf r} = (\rho, y)$
(in cylindrical coordinates $\rho, \varphi, y$). For a homogeneous
applied field the applied vector potential in these two geometries
reads  $A_a = -x B_a$ or $A_a = -\rho B_a /2$. The current density
${\bf J(r},t)$, electric field ${\bf E(r},t)$, and vector
potential ${\bf A(r},t)$ now have only one component oriented
along $z$ or $\varphi$ and denoted by $J$, $E$, $A$. The
method \cite{15,16,18}  describes the superconductor by its
current density $J({\bf r},t)$, from which the magnetic field
${\bf B}(x,y,t) =(B_x, B_y)$ or ${\bf B}(\rho,y,t) =(B_\rho, B_y)$,
the magnetic moment $m(t)$ (along $y$), and the electric field
$E({\bf r},t) =E(J,{\bf B,r'})$  follow directly or via the
constitutive law $E=E(J,{\bf B})$. For high inductions
$B \gg \mu_0 H_{c1}$ one has  ${\bf B} = \mu_0 {\bf H}$ everywhere
and $J = -\nabla^2 (A-A_a)$. The current density $J$ is then
obtained by time-integrating the following equation of motion,
  \begin{eqnarray}   
  \dot J({\bf r},t) = \int_V \! {\rm d}^2 r'\, K({\bf r,r'})\, [\,
  E(J,{\bf B}) +  \dot A_a({\bf r'},t) \,]\,.
  \end{eqnarray}
Here $K({\bf r,r'}) = Q({\bf r,r'})^{-1}$ is an inverse integral
kernel obtained by inverting a matrix, see \cite{15,16} for details.
The kernels $Q$ and $K$ apply to the appropriate geometry and
relate $J$ to the current-caused vector potential
$A-A_a$ in the (here trivial) gauge  $\nabla\cdot {\bf A}=0$
via integrals over the specimen volume $V$,
  \begin{eqnarray}    
  A({\bf r}) &=& -\!\int_V \! {\rm d}^2r'\,Q({\bf r,r'})\,J({\bf r'})
              + A_a({\bf r}) \,,  \\
  J({\bf r}) &=& -\!\int_V \! {\rm d}^2r'\,K({\bf r,r'})\,[A({\bf r'})
              - A_a({\bf r'}) ] \,.
  \end{eqnarray}
The Laplacian kernel $Q$ is universal, e.g.,
$Q({\bf r,r'}) = -(\mu_0 /2\pi) \ln|{\bf r-r'}|$ for long strips
with arbitrary cross section, but the inverse kernel $K$
depends on the shape of the specimen cross section. Putting
$A({\bf r'}) = 0$ in Eq.\ (10) (Meissner state) one sees that
  \begin{eqnarray}    
  J_s({\bf r}) = \!\int_V \!{\rm d}^2r'\, K({\bf r,r'})\,A_a({\bf r'})
  \end{eqnarray}
is the surface screening current caused by the applied field.
In particular, one has $J_s({\bf r}) = 0$ inside the superconductor.
In our method $J_s$ automatically is restricted to the layer
of grid points nearest to the surface, see Fig.\ 2.

   If one is interested also in low inductions one has to generalize
Eq.\ (8) to general reversible magnetization ${\bf H=H(B)}$. This is
achieved by replacing in the constitutive law ${\bf E(J,B)}$ the
genuine current density ${\bf J}=\mu_0^{-1} \nabla \times {\bf B}$ by
the effective current density ${\bf J_H = \nabla \times H}$ which
drives the vortices and thereby generates an electric field ${\bf E}$.
That ${\bf J_H = \nabla \times H(B,r)}$ enters the Lorentz
force is rigorously proven by Labusch \cite{17}. Within the London
theory this important relation may also be concluded from the
facts that the force on a vortex is determined by the {\it local}
current density at the vortex center, while the energy density
$F$ of the vortex lattice is  deter-
mined by the magnetic field at the vortex centers. Thus,
${\bf J_H} = \nabla \times (\partial F/\partial {\bf B})$
is the average of the current densities at the vortex centers,
which in general is different from
\linebreak \newpage

  \begin{figure}[F5]
\epsfxsize= .95\hsize  \vskip 1.5\baselineskip
\centerline{ \epsffile{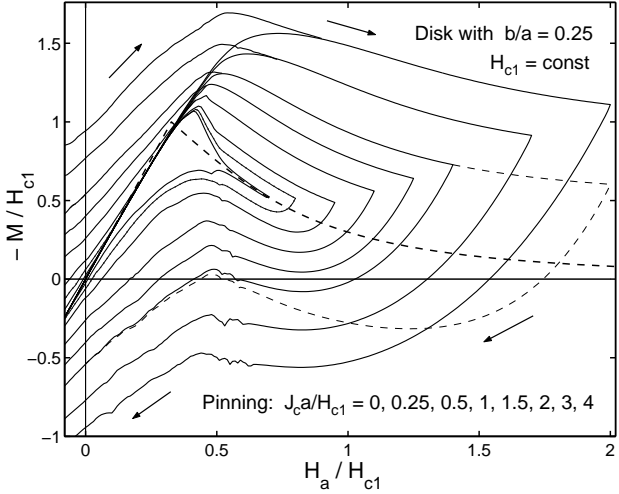}}
\caption{The magnetization curves $M(-H_a)= -M(H_a)$ of a thick disk
 with aspect ratio  $b/a=0.25$ and constant $H_{c1}$ for various
 pinning strengths,  $J_c$ = 0, 0.25, 0.5, 1, 1.5, 2, 3, 4 in units
 $H_{c1}/a$, and various sweep amplitudes. Bean model.
 The inner loop belongs to the pin-free disk ($J_c=0$), the outer
 loop to strongest pinning. The reversible  magnetization curve of
 the corresponding ellipsoid is shown as a dashed curve.
  } \end{figure}

\noindent
the current density
${\bf J} = \mu_0^{-1}\nabla \times {\bf B}$ averaged over the
vortex cells. In our 2D geometry one thus has to replace in
Eq.\ (8)
  \begin{eqnarray}    
  E[\,   J({\bf r}'), {\bf B(r'}) \,]  \to
  E[\, J_H({\bf r}'), {\bf B(r'}) \,] \,,
  \end{eqnarray}
where $J_H = \partial H_y /\partial x - \partial H_x /\partial y$
depends on the reversible material law
$H(B)=\partial F /\partial B$ with  $H_x = H(B) B_x/B$,
$H_y = H(B) B_y/B$, and $B= (B_x^2 +B_y^2)^{1/2}$.

   The boundary condition on ${\bf H(r)}$ is simply that
one has ${\bf H = B}/\mu_0$ at the surface (and in the vacuum
outside the superconductor, which does not enter our calculation).
This boundary condition may be forced by an appropriate
space-dependent material law ${\bf H = H(B,r)}$, which
outside and at the surface of the superconductor is trivially
${\bf H = B}/ \mu_0$. The specimen shape thus enters in two
places: via the integral kernel
$K({\bf r,r'})$ and via the material law ${\bf H=H(B,r)}$.

  To compute the induction ${\bf B(r)}$ entering ${\bf H(B)}$,
for maximum accuracy one should not use the derivative
${\bf B = \nabla \times A}$ but the Biot-Savart integral
  \begin{eqnarray}    
  {\bf B(r)} = \!\int_V \! {\rm d}^2 r'\, {\bf L(r,r')} J(r')
              + {\bf B_a(r)} \,,
  \end{eqnarray}
with appropriate kernel ${\bf L(r,r')}$. The accuracy of the
method then depends mainly on the algorithm used to compute the
derivative ${\bf J_H = \nabla \times H}$. A useful trick is to
compute ${\bf J_H}$ as ${\bf J_H =J + \nabla \times (H-B}/\mu_0)$
where ${\bf H-B}/\mu_0$ is typically small and vanishes at the surface.

  \begin{figure}[F6]
\epsfxsize= .95\hsize  \vskip 1.5\baselineskip
\centerline{ \epsffile{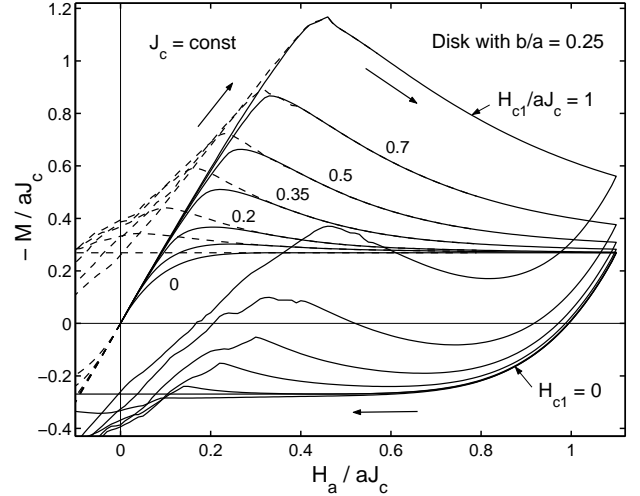}}
\caption{Magnetization curves of a disk as in Fig.\ 5 but
 with $J_c=$const.\ and for various lower critical fields
 $H_{c1}$ = 0, 0.2, 0.35, 0.5, 0.7, 1 in units $aJ_c$.
 Bean model.
  } \end{figure}

  For the following computations I use simple models for the
constitutive laws of an isotropic homogeneous type-II superconductor
without Hall effect, though our method \cite{18} is more general.
With  Eq.\ (6) and $H = B/\mu_0 - M$ one has
  \begin{eqnarray}    
   H(B) = \mu_0^{-1} [ B_{c1}^3 + B^3 ]^{1/3} \,.
  \end{eqnarray}
with $B_{c1} = \mu_0 H_{c1}$. A simple $B$-dependent
current--voltage law which describes pinning, thermal depinning,
and flux flow is ${\bf E(J,B)} = \rho(J,B){\bf J}$ with
  \begin{eqnarray}    
  \rho(J,B) = \rho_0 B {(J/J_c)^\sigma \over 1 + (J/J_c)^\sigma}\,.
  \end{eqnarray}
This model has the correct limits $\rho \propto J^\sigma$
($J\ll J_c$, flux creep) and $\rho =\rho_0 B = \rho_{\rm FF}$
($J\gg J_c$, flux flow, $\rho_0 =$ const),
and for large creep exponent $\sigma \gg 1$ it reduces to the Bean
critical state model. In general the critical current density
$J_c = J_c(B)$ and the creep exponent $\sigma(B) \ge 0$ will
depend on $B$. For pin-free superconductors ($J_c \to 0$) this
expression describes usual flux flow, i.e.,  viscous motion of
vortices, ${\bf E}=\rho_{\rm FF}(B) {\bf J}$,  with flux-flow
resistivity $\rho_{\rm FF} \propto B$ as it should be.

\section{Pin-free superconductors}  

   The penetration and exit of flux computed from Eqs.\ (8-15)
is visualized in Figs.\ 1 -- 3 for isotropic strips and disks
without volume pinning, using a flux-flow resistivity
$\rho_{\rm FF} = \rho_0 B({\bf r})$ with $\rho_0 = 140$ (strip) or
$\rho_0 = 70$ (disk)
in units where $H_{c1} =a=\mu_0=|dH_a/dt| =1$.
Figure 1 shows the field lines of ${\bf B}(x,y)$ in two pin-free
strips with aspect ratios $b/a=2$ and $b/a=0.3$; Fig.\ 2 shows
the surface screening currents in the same strips before flux
has penetrated; and Fig.\ 3 plots some induction profiles
in a strip and some hysteresis loops of the magnetization
\linebreak\newpage

  \begin{figure}[F7]
\epsfxsize= .95\hsize  \vskip 1.5\baselineskip
\centerline{ \epsffile{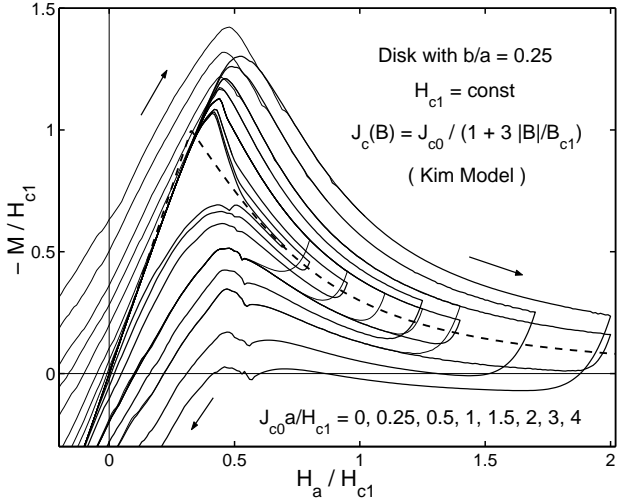}}
\caption{Magnetization curves of the same disk as in Fig.\ 5 but
 for the Kim model, $J_c(B)= J_{c0}/(1+3|B|/B_{c1})$ for various
 pinning strengths  $J_{c0}$ = 0, 0.25, 0.5, 1, 1.5, 2, 3, 4
 in units $H_{c1}/a$. Presentation as in Fig.\ 5.
  } \end{figure}

\noindent
and of the induction in the center of a strip and disk.

 The profiles of the induction $B_y(r,y)$ taken along the midplane
$y=0$ of the thick disk in Fig.\ 3 have a pronounced minimum
near the edge $r=a$, which is the region where strong screening
currents flow. Away from the  edges, the current density
${\bf J} = \nabla \times {\bf B}/\mu_0$ is nearly zero; note
the parallel field lines in Fig.\ 1. The quantity
${\bf J_H} = \nabla\times{\bf H(B)}$  which enters the Lorentz
force density ${\bf J_H \times B}$, is even exactly zero since
we assume absence of pinning and the viscous drag force is small.
Our finite flux-flow parameter $\rho_0$ and finite ramp
rate ${\rm d}H_a / {\rm d}t = \pm 1$ mean a dragging force
which, similar to pinning, causes a weak hysteresis and a small
remanent flux at $H_a =0$; this artefact is reduced by choosing
a larger resistivity or a slower ramp rate.

   In Fig.\ 3 the induction $B_0=B_y(0,0)$ in the specimen
center performs a hysteresis loop very similar to the magnetization
loops $M(H_a)$ shown in Figs.\ 3 and 4. Both loops are symmetric,
$M(-H_a) = -M(H_a)$ and $B_0(-H_a) = -B_0(H_a)$. The maximum of
$M(H_a)$  defines a field of first flux entry $H_{\rm en}$, which
closely coincides with the field $H'_{\rm en}$ at which $B_y(0,0)$
starts to appear. The computed entry fields are well fitted by
       \begin{eqnarray}    
  H_{\rm en}^{\rm strip}/H_{c1}  &=& \tanh \sqrt{0.36 b/a} \,,
                             \nonumber \\
  H_{\rm en}^{\rm disk}/H_{c1}   &=& \tanh \sqrt{0.67 b/a} \,.
       \end{eqnarray}
These formulae are good approximations for all aspect ratios
$0 < b/a < \infty$, see also the estimates of
$H_{\rm en} \approx \sqrt{b/a}$ for thin strips in
Refs.\ \onlinecite{6,10}.

   The virgin curve of the irreversible $M(H_a)$ of strips and
disks at small $H_a$ coincides with the ideal Meissner straight
line $M = -H_a/(1-N)$ of the corresponding ellipsoid, Eqs.\ (4,7).
When the increasing $H_a$ approaches
$H_{\rm en}$, flux starts to penetrate into the corners in form of
\linebreak
  \begin{figure}[F8]
\epsfxsize= .95\hsize  \vskip 1.5\baselineskip
\centerline{ \epsffile{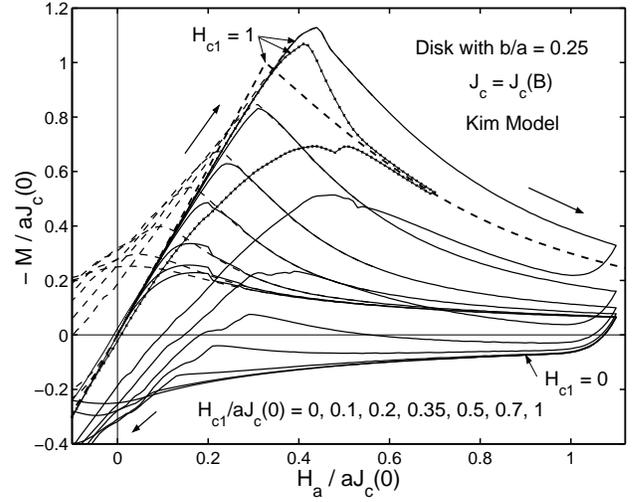}}
\caption{Magnetization curves as in Fig.\ 6 but for the Kim model
 $J_c(B) =J_{c0} /(1 +3|B|/aJ_{c0})$ with $J_{c0}$=const.\  for
 varios $H_{c1}$ = 0, 0.2, 0.35, 0.5, 0.7, 1 in units $aJ_{c0}$.
 Also depicted are the pin-free magnetization (line with dots;
 $M$ and $H_a$ here are in units $H_{c1}$ since $J_{c0}=0$) and
 the irreversible magnetization of the corresponding ellipsoid.
  } \end{figure}
\noindent
almost straight flux lines (Fig.\ 1) and thus $|M(H_a)|$ falls
below the Meissner line.
At $H_a= H_{\rm en}$ flux penetrates and jumps
to the center, and $|M(H_a)|$ starts to decrease. In decreasing
$H_a$, this barrier is absent. As soon as flux exit starts, all
our calculated $M(H_a)$ exhibit strong ``numerical noise'',
which reflects the instability of this state. Similar but weaker
noise occurs at the onset of flux penetration.

  As can be seen in Fig.\ 4, above
some field $H_{\rm rev}$, the magnetization curve $M(H_a)$ becomes
reversible and exactly coincides with the curve of the ellipsoid
defined by Eqs.\ (3, 6, 7) (in the quasistatic limit with
$\rho_0^{-1} {\rm d}H_a/{\rm d}t \to 0$). The irreversibility field
$H_{\rm rev}$ is difficult to compute since it slightly depends on
the choices of the flux-flow parameter $\rho_0$ (or ramp rate) and of
the numerical grid, and also on the model for $M(H_a; 0)$. In the
interval $0.08 \le b/a \le 5$ we find with relative error of $3\%$,
       \begin{eqnarray}    
  H_{\rm rev}^{\rm strip}/H_{c1}  &=& 0.65 +0.12 \ln{(b/a)} \,,
                              \nonumber \\
  H_{\rm rev}^{\rm disk}/H_{\rm c1}   &=& 0.75 +0.15 \ln{(b/a)} \,.
       \end{eqnarray}
This fit obviously does not apply to very small $b/a \ll 1$
(since $H_{\rm rev}$ should exceed $H_{\rm en} > 0$) nor to
very large $b/a \gg 1$ (where $H_{\rm rev}$ should be close
to $H_{\rm c1}$). The limiting value of $H_{\rm rev}$ for thin
films with $b \ll a$ is thus not yet known.

   Remarkably, the irreversible magnetization curves $M(H_a)$ of
pin-free strips and disks fall on top of each other if the strip is
chosen twice as thick as the disk,
$(b/a)_{\rm strip} \approx 2(b/a)_{\rm disk}$. This striking
coincidence holds for all aspect ratios $0 < b/a < \infty$ and can
be seen from
\linebreak\pagebreak

  \begin{figure}[F9]
\epsfxsize= .95\hsize  \vskip 1.5\baselineskip
\centerline{ \epsffile{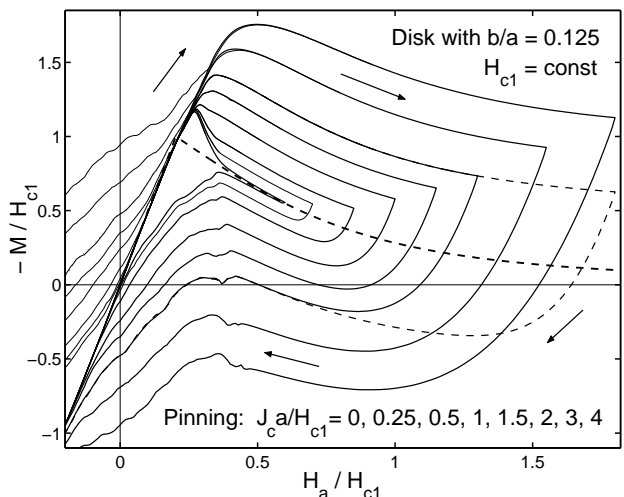}}
\caption{Same magnetization curves as in Fig.\ 5 but for a thinner
 disk with aspect ratio $b/a=0.125$ for various degrees of pinning
 and constant $H_{c1}$.
  } \end{figure}

\noindent
each of Eqs.\ (7,16,17): The effective $N$ [or virgin slope
$1/(1-N)$], the entry field $H_{\rm en}$, and the reversibility field
$H_{\rm rev}$  are nearly equal for strips and disks with half
thickness, or for slabs and cylinders with half length.

   Another interesting feature of the pin-free magnetization loops is
that the maximum of $|M(H_a)|$ exceeds the maximum of the reversible
curve (equal to $H_{c1}$) when $b/a \le 0.8$ for strips and
$b/a \le 0.4$ for disks, but at larger $b/a$ it falls below $H_{c1}$.
The maximum magnetization may be estimated from the slope of the
virgin curve $1/(1-N)$, Eq.\ (7), and from the field of first
flux entry, Eq.\ (16).

   The formulae (7,16,17) are derived essentially from first
principles, with no assumptions but the geometry and finite $H_{c1}$.
They should be used to interpret experiments on superconductors with
no or very weak vortex pinning. At present it is not clear how the
presence of a microscopic Bean-Livingston barrier may modify
these continuum theoretical results.

\section{Superconductors with pinning}  

    Figures 5-8 show how the irreversible magnetization loops
of disks with $b/a=0.25$  (and in Fig.\ 9 for a thinner disk with
$b/a = 0.125$) are modified when volume pinning is switched on.
In Figs.\ 5, 6, and 9, pinning is described by the
Bean model with constant critical current density $J_c$, while
in Figs.\ 7 and 8 the Kim model is used with an induction-dependent
$J_c(B) = J_{c0} / (1 + 3 |B| /B_K)$ with $B_K =\mu_0 H_{c1}/3$
(Fig.\ 8) or $B_K = \mu_0 a J_{c0}/3$ (Fig.\ 9). In Figs.\ 5, 7,
and 9, $H_{c1}$ is held constant; with increasing $J_c$ or $J_{c0}$
(in natural units $H_{c1}/a$) the magnetization loops
are inflated nearly symmetrically about the pin-free loop or
about the reversible curve (above $H_{\rm rev}$), and the maximum
of $|M(H_a)|$ shifts to higher fields.
Above $H_{\rm rev}$ the width of the loop is nearly proportional
to $J_c$, as expected from theories \cite{15,16} which
assumed $H_{c1}=0$, but at small fields the influence of finite
$H_{c1}$ is clearly seen up to rather strong pinning.

 In Figs.\ 6 and 8, $J_c$ or $J_{c0}$ is held constant and
$H_{c1}$ increased from zero (in natural units $aJ_c$).
As expected, the influence of finite $H_{c1}$ is most pronounced
at small applied fields $H_a$, where it causes a peak in $-M$
even in the Bean magnetization curves, which without
consideration of $H_{c1}$ consist of two monotonic branches
and a monotonic virgin curve. Within the Kim model, or with
any decreasing $J_c(B)$ dependence, the magnetization loops
exhibit a maximum even when $H_{c1}=0$ is assumed \cite{19}. With
increasing $H_{c1}$ this maximum becomes sharper and shifts to
larger fields, cf.\ Fig.\ 8. Comparing Figs.\ 5 and 9 one
sees that for superconductor disks with pinning and with
$H_{c1}>0$, the peak in $-M(H_a)$ becomes more pronounced and
shifts towards smaller applied fields when the disk
thickness is decreased.


\end{multicols}

\vspace{-0.6 cm}
\references
    \vspace{-1.75 cm}  
\bibitem{1} P.\ W.\ Anderson,  \prl{\bf 9} (1962) 309.

\bibitem{2} J.\ Provost, E.\ Paumier, and A.\ Fortini,
            J.\ Phys.\ F {\bf 4}, 439 (1974); A.\ Fortini,
            A.\ Haire, and E.\ Paumier, \prb{\bf 21} (1980) 5065.
\bibitem{3} C.\ P.\ Bean and J.\ D.\ Livingston, \prl{\bf 12} (1964)
            14;  L.\ Burlachkov, \prb{\bf 47} (1993) 8056.

\bibitem{4} M.\ V.\ Indenbom, H.\ Kronm\"uller, T.\ W.\ Li,
            P.\ H.\ Kes, and A.\ A.\ Menovsky,
            Physica C {\bf 222} (1994) 203;  
            M.\ V.\ Indenbom and E.\ H.\ Brandt,
            \prl{\bf 73} (1994) 1731;
            E.\ H.\ Brandt, Rep.\ Prog.\ Phys.\ {\bf 58} (1995) 1465.

\bibitem{5}  N.\ Morozov et al., \prl{\bf 76} (1996) 138;
       N.\ Morozov et al., Physica C {\bf 291} (1997) 113;

\bibitem{6} E.\ Zeldov, A.\ I.\ Larkin, V.\ B.\ Geshkenbein,
       M.\ Konczykowski, D.\ Majer, B.\ Khaykovich, V.\ M.\ Vinokur,
       and H.\ Strikhman, \prl{\bf 73} (1994) 1428.

\bibitem{7} E.\ Zeldov et al., Physica C {\bf 235-240} (1994) 2761;
       B.\ Khaykovich et al., Physica C {\bf 235-240} (1994) 2757;

\bibitem{8} I.\ L.\ Maksimov and A.\ A.\ Elistratov,
       Pis'ma Zh.\ Eksp.\ Teor.\ Fiz. {\bf 61} (1995) 204
       [Sov.\ Phys.\ JETP Lett. {\bf 61} (1995) 208];
       I.\ L.\ Maximov and A.\ A.\ Elistratov, Appl.\ Phys.\ Lett.
       {\bf 72} (1998) 1650.

\bibitem{9} Th.\ Schuster, M.\ V.\ Indenbom, H.\ Kuhn, E.\ H.\ Brandt,
       and M.\ Konczykowski, \prl{\bf 73} (1994) 1424.

\bibitem{10} M.\ Benkraouda and J.\ R.\ Clem, \prb{\bf 53} (1996) 5716;
       \prb{\bf 58} (1998) 15103.

\bibitem{11} A.\ V.\ Kuznetsov, D.\ V.\ Eremenko, and V.\ N.\ Trofimov,
       \prb{\bf 56} (1997) 9064; \prb{\bf 57} (1998) 5412.

\bibitem{12} L.~D.~Landau and E.~M.~Lifshitz, {\it Electrodynamics
        of Continuous Media}, Vol.\ 8 of Course in Theoretical
        Physics (Pergamon Press, London, 1959).

\bibitem{13} G.\ P.\ Mikitik and E.\ H.\ Brandt,
       \prb{\bf 60} (1999) 592.

\bibitem{14} E.\ H.\ Brandt, \prl{\bf 78} (19997) 2208.  

\bibitem{15} E.\ H.\ Brandt, \prb{\bf 54} (1996) 4246. 

\bibitem{16} E.\ H.\ Brandt, \prb{\bf 58} (1998) 6506, 6523.

\bibitem{17} R.\ Labusch and T.\ B.\ Doyle, Physica C {\bf 290} (1997)
       143; T.\ B.\ Doyle, R.\ Labusch, and R.\ A.\ Doyle,
       Physica C {\bf 290} (1997) 148.

\bibitem{18} E.\ H.\ Brandt, \prb{\bf 59} (1999) 3369;
             {\bf 60} (1999) 11939.

\bibitem{19} D.\ V.\ Shantsev et al., \prl{\bf 82} (1999) 2947.
\end{document}